\documentclass[pdftex,prb,twocolumn,superscriptaddress,showpacs]{revtex4-1}

\usepackage{subfigure}
\usepackage{wrapfig}
\usepackage{textcomp}
\usepackage{graphicx}
\usepackage{amssymb}
\usepackage{amsmath}
\usepackage{bm}
\usepackage{amsmath, amsthm, amssymb}
\usepackage{color}

\def\beq{\begin{equation}}
\def\eeq{\end{equation}}
\def\bea{\begin{eqnarray}}
\def\eea{\end{eqnarray}}

\begin{document}

\title{Stability of edge states in strained graphene}

\author{Pouyan Ghaemi}
\affiliation{Department of Physics, University of Illinois at Urbana-Champaign, Urbana, IL 61801, USA}

\author{Sarang Gopalakrishnan}
\affiliation{Department of Physics, University of Illinois at Urbana-Champaign, Urbana, IL 61801, USA}
\affiliation{Department of Physics, Harvard University, Cambridge, MA 02138, USA}

\author{Shinsei Ryu}
\affiliation{Department of Physics, University of Illinois at Urbana-Champaign, Urbana, IL 61801, USA}

\begin{abstract}
Spatially inhomogeneous strains in graphene can simulate the effects of valley-dependent magnetic fields. As demonstrated in recent experiments~\cite{levy2010, manoharan}, the realizable magnetic fields are large enough to give rise to well-defined flat pseudo-Landau levels, potentially having counter-propagating edge modes. In the present work we address the conditions under which such edge modes are visible. We find that, whereas armchair edges do not support counter-propagating edge modes, zigzag edges do so, through a novel selective-hybridization mechanism. We then discuss effects of interactions on the stability of counter-propagating edge modes, and find that, for the experimentally relevant case of Coulomb interactions, interactions typically decrease the stability of the edge modes. Finally, we generalize our analysis to address the case of spontaneous valley polarization, which is expected to occur in charge-neutral strained graphene~\cite{pouyan12, pesin}. 
\end{abstract}

\maketitle

\section{Introduction}

Among condensed-matter systems, graphene is unique in being a flexible two-dimensional membrane whose electronic properties are tunable through deformations and strains. Spatially varying deformations affect the band structure of graphene by modulating the hopping amplitudes between lattice sites. Indeed, certain spatially varying deformation patterns can mimic the effects of uniform ``pseudo-magnetic fields,'' which have opposite signs in the two low-energy valleys \cite{guinea} (i.e., in the neighborhoods of the $\mathbf{K}$ and $\mathbf{K}'$ Dirac points) of graphene. (Unlike real magnetic fields, therefore, pseudo-magnetic fields preserve overall time-reversal invariance.) To date, such pseudo-magnetic fields have been realized using two distinct experimental approaches~\cite{levy2010, manoharan}; in both experiments, the fields realized were strong enough to drive the electronic structure deep into the quantum Hall regime in each valley, and---in line with theoretical predictions---the electronic structure measured by scanning-tunneling microscopy (STM) was clearly seen to consist of well-spaced pseudo-Landau levels (PLLs) in each valley. While the PLLs in each valley, considered separately, have nontrivial topological invariants~\cite{pouyan12,bitan}, these are opposite for the two PLLs in the $\mathbf{K}$ and $\mathbf{K}'$ valleys; thus, the fate of various topological features in the \textit{full} system is not completely understood.

In the present work, we address the nature of one of the crucial topological features of strained graphene, namely its \textit{edge states}. We first show, by analyzing the noninteracting system, that the Landau levels of strained graphene are very sensitive to the distinction between zigzag and armchair edges. For zigzag edges, the Landau levels (though flat in an infinite system) acquire an appreciable linear dispersion due to their hybridization with the pre-existing, non-topological ``surface'' states. (We shall use this term to distinguish these states from the topological ``edge'' states.) On the other hand, for armchair edges, where such boundary states do not exist, the Landau levels do not disperse even near the edge; this is in sharp contrast with the case of a real magnetic field~\cite{dima1,dima2}; as we discuss, the physics behind this difference is that the lowest PLLs in the two valleys of strained graphene live on the same sublattice, whereas in the presence of a real magnetic field the corresponding Landau levels live on \textit{opposite} sublattices. Thus, on a typical sample with rough edges, the edge states are expected to be localized in the zigzag regions and as a consequence should not contribute to transport; however, two-terminal measurements on zigzag nanoribbons should reveal the presence of multiple edge states. We then address the possibility that Luttinger-liquid effects stabilize the edge states against disorder, but find that they do not for the case of a Coulomb interaction. Finally, we generalize our analysis to the case of edge states in spontaneously valley-polarized states, and argue that, in general, these should not exhibit protected edge states. (One further mechanism for the destabilization of edge states, for instance in the experiments of Ref.~\onlinecite{levy2010}, is their hybridization with
the Dirac sea in the surrounding, non-strained region; however, for undoped or weakly doped graphene this hybridization is presumably weak, owing to the vanishing density of states at the Dirac point.)

Our paper is organized as follows. In Sec.~\ref{sec:model} we introduce the microscopic Hamiltonian for strained graphene that was used in our numerical work, as well as an effective low-energy description that enables us to arrive at an analytic understanding of the effects discussed here. We then consider the physics of an armchair edge in Sec.~\ref{sec:armchair} and that of a zigzag edge in Sec.~\ref{sec:zigzag}. In Sec.~\ref{sec:glazman} we extend our analysis to include interaction effects and their influence on edge stability. In Sec.~\ref{sec:haldane} we generalize our arguments, in the specific case of an armchair edge, to the case of a valley-polarized state. Finally, Sec.~\ref{sec:conclusions} summarizes our results and discusses possible experimental signatures of edge-state physics in strained graphene.

\section{Model and bulk properties} \label{sec:model}

\begin{figure}
\begin{center}
\includegraphics[width=3.3in, height = 3in]{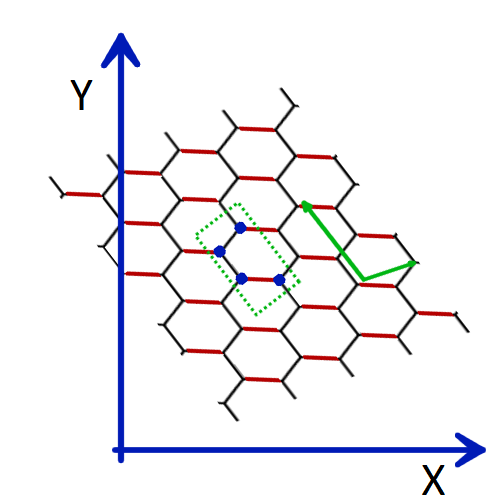}
\caption{Schematic pattern of change in hopping matrix element on red bonds and varying along $x$ direction as $\delta t_3 =e v_F B x$  which leads to the pseudo-vector potential $A_x=0$ and $A_y=Bx$. The dotted green square is the corresponding four cite unite cell and the green arrows are primitive lattice vectors.}
\label{strainpattern}
\end{center}
\end{figure}

In what follows, we shall chiefly consider the following noninteracting, nearest-neighbor Hamiltonian for strained graphene:
\begin{equation}
H_0 =\sum_{\mathbf{r}_{i}} \sum_{a=1,2,3} (t+\delta t_{a} (\bold{r}_{i})) ( a^\dagger (\bold{r}_{i}) b (\bold{r}_{i} + \mbox{\boldmath$\delta$}_a) + h.c. ),
\label{hamiltonian1}
\end{equation}
where $\delta t_{a} (\bold{r}_{i})$ is, the strain-induced, nearest neighbour hopping amplitude
modulation between the $A$-sublattice site at $\bold{r}_{i} $ and the $B$-sublattice site at $\bold{r}_{i} + \mbox{\boldmath$\delta$}_a$ of the bipartite honeycomb lattice \cite{mpr}. The bond $\mbox{\boldmath$\delta$}_{a}$ connects any $A$-sublattice atom to its three nearest neighbors on the $B$-sublattice. In the absence of strain, the low-energy excitations have a linear dispersion around the two Dirac points at momenta $\pm {\bf K}$ with ${\bf K}=(4 \pi/3 \sqrt{3}a_0) {\bf e_x}$, $a_0$ being the carbon-carbon bond length \cite{mpr}.
Near the Dirac points $\mathbf{K}$ and $\mathbf{K}'$ one can
write the Bloch states in terms of
a four-component spinor, as follows: 
$\Psi \equiv (\psi_{A,\mathbf{K}}, \psi_{B,\mathbf{K}}, \psi_{A,\mathbf{K}'}, \psi_{B,\mathbf{K}'})$ 
where the first index denotes the component of the wavefunction on the $A(B)$ sublattice of the honeycomb unit cell, and the second index denotes the component of the state that is associated with the $\mathbf{K}$ ($\mathbf{K}'$) valley. (In what follows, we shall write $\psi_{A, \mathbf{K(K')}} \equiv u_{\mathbf{K(K')}}$ and $\psi_{B, \mathbf{K(K')}} \equiv v_{\mathbf{K(K')}}$, in order to make contact with the standard notation for Dirac fermions.) The low energy effective Hamiltonian close to the Dirac points then reads as: 
\beq
\label{kn}
\mathcal{H}_0 = v_F \left[\hat{p}_x \Gamma_x + \hat{p}_y \Gamma_y\right]
\eeq
where $\Gamma_x =\tau_3 \sigma_1$, $\Gamma_y = \tau_0 \sigma_2$, $v_F$ is the Fermi velocity, and the $\sigma$ and $\tau$ operators are Pauli matrices acting on sublattice and valley indices respectively. 

Strain generates a
pseudo vector potential given by $A_x^0+i A_y^3=\sum_{a=1,2,3} \delta t_{a}(\textbf{r})e^{\pm i \textbf{K} \cdot \mbox{\boldmath$\delta$}_{a}}$ near the Dirac points $\pm \mathbf{K}$\cite{guinea,mpr}.
 Note that $\sum_{a=1,2,3} \delta t_{a}(\textbf{r})e^{\pm i \textbf{K} \cdot \mbox{\boldmath$\delta$}_{a}}$ is complex because the nearest-neighbor hoppings are not symmetric under inversion. The real part of the strain gauge field $ A_x^0$ is the same in both valleys and so couples to  $Q_0=\sigma_0\tau_0$ and can be gauged away assuming time-reversal symmetry holds; whereas the imaginary part $i A_y^3$, has opposite sign in the two valleys and couples with\cite{us:su2} $Q_3=\sigma_0\tau_3$  leading  to the valley-dependent magnetic fields realized in the experiments of Ref.~\onlinecite{levy2010}. 
 
 For the purposes of the present work, we shall consider ribbon geometries, in which the strain-induced field is taken to realize
the Landau ``gauge'' ($\mathbf{A} = (0, Bx)$).
 A concrete lattice realization of this gauge field is shown in Fig.~\ref{strainpattern}. 
In this realization of strain,
 exploiting translation invariance in $y$ direction, the Dirac Hamiltonian can be written in the form:
\begin{equation}
\label{kn}
\mathcal{H}_0 = v_F \left[-i\partial_x \tau_3 \sigma_1 +\hat{p}_y\tau_0\sigma_2-\frac{e}{c} B x \tau_3\sigma_2 \right]
\end{equation}
where $v_F$ is the Fermi velocity, and the $\sigma$ and $\tau$ operators are Pauli matrices acting on sublattice ($u$ and $v$)  and valley ($\textbf{K}$ and $\textbf{K}'$) indices respectively.
As this strain pattern respects translation invariance in the $y$ direction, it is suitable for a strained ribbon extended in the $y$ direction. Notice that the Hamiltonian for valley $\textbf{K}$ associated with $p_y$ is the same as the Hamiltonian for valley $\textbf{K}'$ with momentum $-p_y$.

In this Landau ``gauge'',
the wavefunctions in the zeroth PLL have the four-component form   $\Psi^{\mathbf{K}}_{0,p_y} = (\phi_{0,p_y},0,0,0)$ 
and 
$\Psi^{\mathbf{K}'}_{0,p_y} = (0,0,\phi_{0,p_y}^*,0)$,
respectively, 
where $\phi_{0,p_y}$ is the $m$th Landau orbital in the lowest 
(nonrelativistic) Landau level, namely $\phi_{p_y}\propto \exp(-ip_y y - (x - p_y l_M^2)^2/2 l_M^2)$ where $l_M$ is the magnetic length associated with the strain induced pseudo-magnetic field. In the higher PLLs, the wavefunctions take the form $\Psi^{\mathbf{K}}_{n,p_y} = (\phi_{n,p_y},\phi_{n-1,p_y},0,0)$ and $\Psi^{\mathbf{K}}_{n,p_y} = (0,0,\phi_{n,p_y}^*,\phi_{n-1,p_y}^*)$. These forms should be contrasted with those for graphene in a real magnetic field; in the case of a real field, the Landau level wavefunctions in the two valleys have \textit{opposite} sublattice structure, whereas in the case of a strain-induced field, the wavefunctions have the \textit{same} sublattice structure.  
 
In our treatment of the noninteracting problem we have suppressed the physical spin, and thus ignored the (weak) intrinsic spin-orbit coupling in graphene. However, the spin index can be included trivially. We shall consider the consequences of the physical spin in Sec.~\ref{sec:glazman}, as interaction effects in the spinless and spinful cases are different; moreover, a combination of interaction effects and spin-orbit coupling was argued to lead to stabilization of fractional phases~\cite{pesin}. (The spinless situation can be experimentally realized by applying a large field parallel to the graphene sheet, and thus spin-polarizing the electrons.)  

\section{Armchair edge} \label{sec:armchair}





Having discussed the bulk properties, we now turn to the case of a graphene ribbon with armchair edges along the $y$ direction. Owing to the translational invariance along $y$ as well as our choice of
the Landau ``gauge''
the momentum in $y$ direction is a good quantum number. On the other hand, in the ribbon geometry, translation invariance in $x$ direction is broken. For our analytic calculations we consider the case of a semi-infinite graphene ribbon covering the region $x < 0$, with an armchair boundary at $x = 0$.

\begin{figure}[htp]
\vspace{0.05in}\subfigure[]{\label{sym}
\includegraphics[trim=2cm 5cm 1cm 5cm, clip=true, height=6cm,width=8cm]{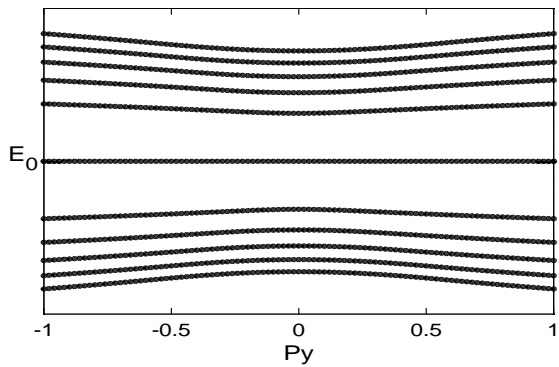} \vspace{0in}}
\subfigure[]{\label{nsym}
\includegraphics[trim=2cm 5cm 1cm 5cm, clip=true, height=6cm,width=8cm]{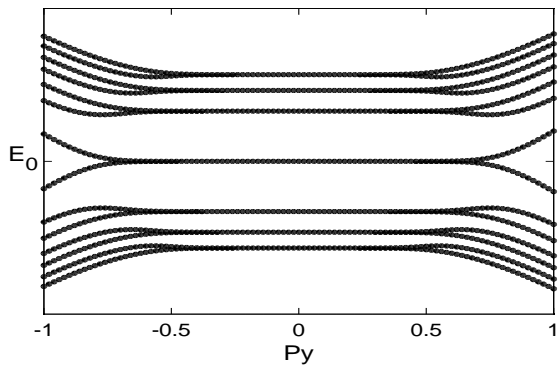}
\vspace{0.05in}}
\caption{
\label{armchair}
Energy dispersion for a graphene nanoribbon with armchair edges
 as a function of the momentum along the edges ($p_y$);
 (a) in the presence of a strain-induced pseudo magnetic field
 and (b)
 in the presence of a real magnetic field.
 The center of mass position of the Landau orbitals
 is correlated with their momentum $p_y$ i.e. $p_y=0$ is localized in the center of the ribbon and as $|p_y|$ increases, they get closer to the edge. 
 For the strain induced magnetic field,
 there is no dispersion in the zeroth PLL,
 whereas for real magnetic field
 there are dispersing edge states associated to
 the topological character of the bulk
 (i.e., the quantum Hall effect). 
} 
\end{figure}

If we make the transformations $ u_{\mathbf{K}'}\rightarrow -u_{\mathbf{K}'}$ and $v_{\mathbf{K}'} \rightarrow -v_{\mathbf{K}'}$, the Hamiltonian in the two valleys reads as:
\begin{eqnarray}
H_\mathbf{K}&=&v_F \left[- i\partial_x \sigma_1+ p_y \sigma_2 -\frac{e}{c} B x \sigma_2\right], \\
H_{\mathbf{K}'}&=&v_F \left[- i\partial_x \sigma_1- p_y \sigma_2 -\frac{e}{c} B x \sigma_2\right].
\end{eqnarray}
In order to get the spectrum, we square these Hamiltonians, leading to:
\begin{eqnarray}
v_F^2\left[-\partial^2_x + \left(p_y-\frac{e}{c}B x\right)^2-\frac{e}{c}B \sigma_3\right]\Psi_k &=& E^2 \Psi_k, \label{k}\\
v_F^2\left[-\partial^2_x + \left(p_y+\frac{e}{c}B x\right)^2-\frac{e}{c}B \sigma_3\right]\Psi_{k'} &=& E^2 \Psi_{k'}, \label{kk}
\end{eqnarray}
which is defined for $x<0$. In the zeroth Landau level, in which (for a strain-induced field) the wave functions in both valleys are based on the same sublattice, so that the $\sigma$ matrix is trivial. The boundary condition then reduces to the wave function vanishing on the last row. In terms of the sublattice wave functions this reads as:
\begin{equation}
u_{\mathbf{K}}(p_y)=u_{\mathbf{K}'}(p_y)
\end{equation}
If we change $x\rightarrow -x$ for $\mathbf{K}'$,
the two equations (\ref{k}) and (\ref{kk}) become identical. We can join the two equations and the boundary condition above will then correspond to continuity of the wave function at $x=0$. The Hamiltonian then reduces to the simple harmonic oscillator Hamiltonian and the boundary question is trivially satisfied. The salient property of this Hamiltonian is the fact that the energy eigenvalues are independent of $p_y$, and thus the states near the edge do not disperse. This is in sharp contrast with the case of a regular magnetic field~\cite{dima1,dima2}, in which states near the edge do disperse. 

One can understand the nondispersing nature of the edge states intuitively as follows, for the case of a semi-infinite ribbon with an edge at $x = 0$. Because both the $(K, k_y)$ and $(K', k_y)$ wavefunctions are located on the same sublattice, and their guiding centers are at $k_y$ and $-k_y$ respectively. One can always construct equal-weight superpositions of the two wavefunctions, which automatically satisfy the boundary condition. (By contrast, in the case of a regular magnetic field, the two states live on opposite sublattices and therefore cannot be superposed to meet the boundary conditions; instead, the boundary condition must be satisfied by introducing a slowly varying envelope function~\cite{dima1,dima2}, and this costs additional kinetic energy for states near the edge.) Note that these arguments extend straightforwardly to the case of higher PLLs. 

As Fig.~\ref{armchair}(a) shows, a numerical computation of the band structure is consistent with the argument above. An important implication of this argument is that there should be no counter-propagating edge modes on armchair edges; i.e., if the chemical potential lies between two PLLs, it should not cross any states at all.




\section{Zigzag edge} \label{sec:zigzag}

We now turn to the case of nanoribbons with zigzag edges. Even in the absence of a magnetic field, these edges support non-dispersing modes localized at the edges of the sample; these ``surface'' states occupy the $A$ sublattice at one end of the system, and the $B$ sublattice at the other end. Because these states are confined on scales smaller than the magnetic length, they are expected to be unaffected by a magnetic or strain field~\cite{dima1,dima2}. In the case of a real magnetic field, the dispersion of the topological edge states~\cite{dima1,dima2} occurs chiefly by intrinsic, universal means, and is therefore only slightly modified by the presence of zigzag surface states. The case of a strain-induced magnetic field is fundamentally different in this respect: as we have argued above, there is no intrinsic dispersion mechanism, and thus the \textit{primary} contribution to the Landau level dispersion is the differential hybridization between Landau levels and the zigzag edge (or ``surface'') states.

\begin{figure}[htp]
\vspace{0.05in}\subfigure[]{\label{sym}
\includegraphics[trim=2cm 5cm 1cm 4cm, clip=true, height=7cm,width=8cm]{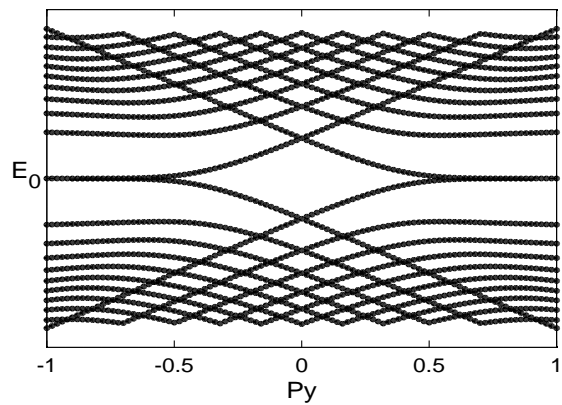} \vspace{0in}}
\subfigure[]{
\includegraphics[trim=2cm 5cm 1cm 5cm, clip=true, height=7cm,width=8cm]{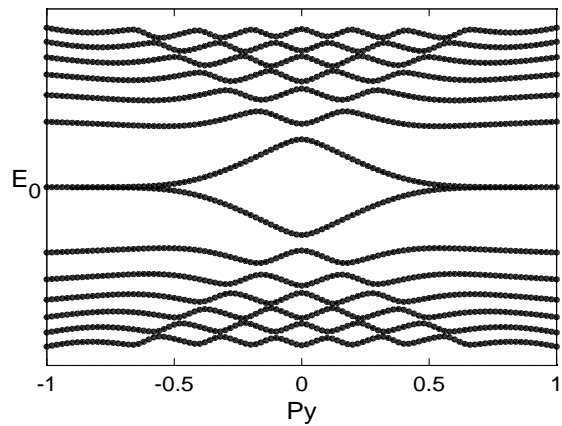}
\vspace{0.05in} \label{nosys}}
\caption{
Energy dispersion for a graphene nanoribbon with zigzag edges
 as a function of the momentum along the edges ($p_y$);
 (a) in the presence of inversion symmetry
 and (b) in the presence of a inversion symmetry breaking perturbation.}
\end{figure}


We now address the effects of this hybridization, focusing at first on the lowest PLL. In the lowest PLL, all Landau orbitals are situated on either the $A$ or the $B$ sublattice. Thus, they hybridize only with the zigzag surface states at one edge of the system; the strength of this hybridization depends on the distance of the guiding center of the Landau orbital from the ``Hybridizing'' edge.

Assuming the hybridizing edge is located at $x = 0$, this hybridization changes the energy of each Landau orbital by a factor $\sim \exp[- (p_y l_M)^2]$.
Because this energy shift is larger for levels situated near the edge, it causes these states to disperse. Moreover, because of the valley-dependence of the pseudo-magnetic field, the states in the $\mathbf{K}$ valley that are near the edge have \textit{opposite} momenta to the corresponding states in the $\mathbf{K}'$ valley. Near the ``non-hybridized'' edge, by contrast, there is no hybridization and thus no dispersion at all. The size of this effect is quite strong in mesoscopic systems (such as all existing experimental realizations of strain-dependent gauge fields), as one can see from diagonalization on graphene nanoribbons (Fig.~\ref{sym}). 

Two important observations can be made about this surface-induced dispersion effect. 

  (1) The hybridization with the zigzag edge state 
  plays a role 
  similar to an electric field perpendicular to the edges of the ribbon;
thus, the dispersion generated by the zigzag surface states can presumably be either enhanced or compensated by the application of such an electric field. In particular, the number of surface states crossing the chemical potential might be detectable by these means. (2) Unlike armchair edges, zigzag edges with surface states do not mix the valleys. This can easily be seen from Fig.~\ref{sym}---the level crossings between PLLs from opposite valleys are \textit{not} avoided, as they would be in the presence of valley mixing. The physical reason for this is that the Landau orbitals in the two valleys---since they are counter-propagating---couple to orthogonal, counter-propagating linear combinations of the zigzag edge states. Thus, for a clean zigzag edge, there should be counterpropagating edge states, which can be detected (e.g.) in two-terminal transport measurements. Note that these counterpropagating edge states can be gapped out by adding a valley-mixing term such as the Kekul\'{e} distortion (see Fig.~\ref{nosys}). 


In practice, perhaps the most promising candidate for realizing a system with a clean zigzag edge is molecular graphene. In this system, one can tunably create a Kekul\'{e} distortion~\cite{manoharan} and thus study the transition between Fig.~\ref{sym} and Fig.~\ref{nosys}.


\section{Stability under interactions}\label{sec:glazman}

The considerations discussed so far apply to the case of noninteracting edge modes. We now address the issue of whether the edge modes continue to carry current in a disordered interacting system, using the approach of Ref.~\cite{glazman}. (Similar questions were addressed for quantum spin-Hall insulators with multiple edge modes in Ref.~\onlinecite{xu:moore}.) It is clear that, in the absence of interactions, backscattering due to disorder localizes the edge modes, leading to an insulating edge. However, there are certain regimes~\cite{xu:moore, glazman, nagaosa} in which interactions prevent the localization of edge states, by making backscattering processes ``irrelevant'' in the renormalization-group sense. (Qualitatively, this effect can be understood in terms of interactions screening out the impurity potential; this would happen, e.g., if the electron-electron interactions were attractive~\cite{kane:fisher}.) 

For repulsive interactions, as discussed in Ref.~\onlinecite{glazman}, the low-temperature behavior of the conductance is given by the strength of the density-density interaction between the left-moving and right-moving edge states (i.e., forward-scattering) compared with back-scattering processes involving momentum-exchange. For the conductance to stay finite as $T \rightarrow 0$, the above criterion implies that
the Fourier components of 
the interaction potential satisfy $V(0) \leq \frac{1}{2} V(2 |\mathbf{G}|)$, 
where $\mathbf{G}$ is a reciprocal lattice vector. (This is an approximate expression; the precise momentum transfer involved in backscattering is the momentum difference between the two states on the edge, which is $\sim |\mathbf{G}| + 2 k_F$, which we approximate as $|\mathbf{G}|$ using the assumption that the chemical potential is near the Dirac point.) This criterion is not met by realistic interaction potentials. However, it has been shown~\cite{glazman} that for $V(0) \leq 2 V(2 |\mathbf{G}|)$ (as in, e.g., a contact potential, or a sufficiently short-ranged potential achieved via gating), the conductance exhibits non-monotonic behavior and increases with decreasing temperature until a crossover scale 

\beq
T^* \sim D \exp\left[ - \frac{ 2 V(2 |\mathbf{G}|) - V(0)}{V(2 |\mathbf{G}|) (2 V(0) - V(2 |\mathbf{G}|)) } \right],
\eeq
where $D$ is the bandwidth of the Dirac band (for molecular graphene this is $\sim 200$ meV). The physics of this crossover can be explored in strained graphene by tuning the range of the inter-electron interaction via gating. 

\section{Effects of valley polarization}\label{sec:haldane}

Thus far, we have considered the edge states of strained graphene under the assumption that time-reversal invariance is preserved. We now turn to situations in which time-reversal invariance is spontaneously \textit{broken}; such situations naturally arise in the case of half-filled PLLs~\cite{pouyan12}, in which the valley-polarized state naively appears to be a topologically protected state. We argue in this section that such states do \textit{not} generically have topologically protected edge modes. 

Within the low-energy theory, there are two valley-polarizing perturbations, given by $\tau_z \sigma_0$ (i.e., simple valley-polarization) and $\tau_z \sigma_z$ (i.e., the Haldane mass term\cite{haldane}). The former opens up a gap in any half-filled PLL, whereas the latter opens up a gap only in the zeroth PLL. We shall consider these terms separately, beginning with the simple valley-polarization term. As discussed in Ref.~\onlinecite{us:su2}, this term anticommutes with two charge-density-wave perturbations that take the form $\tau_x \sigma_0$ and $\tau_y \sigma_0$ in the low-energy theory. Being spatial modulations, these terms are naturally generated at sharp edges; thus, the valley polarization gap in the bulk can be rotated into the valley-mixing gap at the edge without closing any gaps, thus leading to the absence of edge states. We only consider the case of uniform charge-density-wave terms on the edge, and neglect the question of whether disordered or local charge-density-wave terms can gap out the edges. One expects this mechanism to prevent edge states from being present at half-filling of any PLL, as long as (a) the edges mix the valleys, or (b) the charge-density-wave is dynamically generated by interactions at the edge. It is of course possible that very smooth edges, which do not mix the valleys, can support chiral edge states, but the fragility toward valley-mixing indicates that these edge states are not in fact protected. 

\begin{figure}
\begin{center}
\includegraphics[width=3.3in, height = 2in]{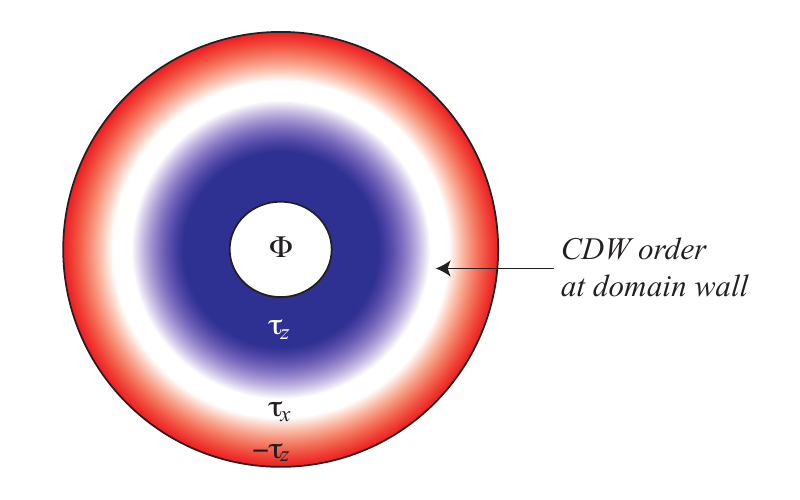}
\caption{Corbino disk with broken valley degeneracy}
\label{corb}
\end{center}
\end{figure}

Before we proceed to the case of the zeroth PLL, we briefly explain how our remarks above can be related to the conventional flux-insertion argument\cite{laghlin}, which was used~\cite{pesin} to argue for topologically protected edge modes in half-filled PLLs. Consider half-filled PLLs in a Corbino disk geometry as shown in Fig. \ref{corb}, in which the left half of the disk has a perturbation of the form $\tau_z$ and the right half has a perturbation of the form $-\tau_z$; this situation, which corresponds to magnetic domains, should typically arise in experiments, given the Ising nature of the valley polarization~\cite{pesin}. We assume that the gauge realized by the strain has the property that 
$p_y=0$ is
near the center of the disk, and then add a perturbation of the form $\tau_x$ in the medial region of the Corbino disk. (This assumption can easily be relaxed, if one then chooses a perturbation that is modulated along $y$.) Now suppose one inserts a (real) magnetic flux through the center of the Corbino disk: if the regions having a $\tau_z$ perturbation were indeed topological, one would expect the flux insertion to pump two charges from the edges of the disk to its center (given the opposite nature of the strain-induced magnetic field in the two regions). However, this is impossible because the center of the disk is fully gapped and therefore incompressible; thus, we conclude that the flux insertion argument fails, and that half-filled PLLs do not in general have a quantized Hall conductance.

\begin{figure}[htbp]
\begin{center}
\includegraphics[trim=3cm 5cm 1cm 4cm, clip=true, height=10cm,width=9cm]{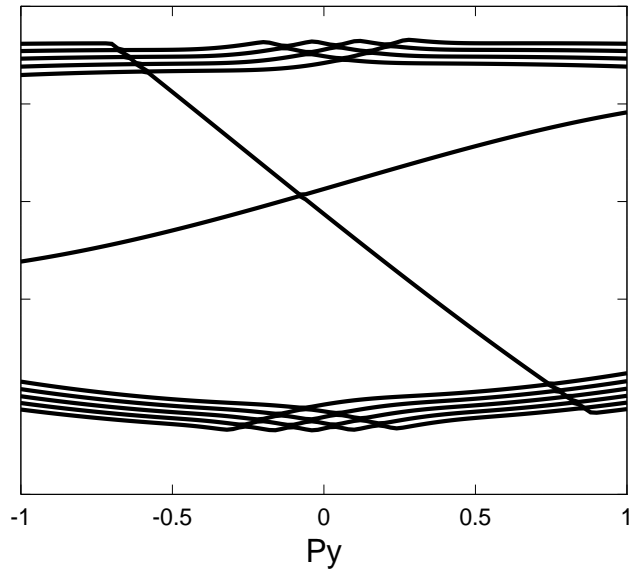} 
\caption{Pseudo-Landau level (PLL) structure in the presence of a large Haldane mass gap. As one can see, the topologically protected edge modes guaranteed by the Haldane mass term do not mix with the PLLs, but appear to originate at much higher energies.}
\label{zerothhaldane}
\end{center}
\end{figure}

The argument given above holds for all the PLLs; however, an additional subtlety arises in the case of the zeroth PLL, owing to the possibility of
a Haldane mass gap, $\tau_z \sigma_z$ \cite{haldane}.
In any PLL other than the zeroth, this term does not open a gap. However, within the zeroth PLL, it does open up a gap, which leads to the following apparent paradox. As the Haldane mass term generates a gap both with and without the strain-induced field, one can imagine adding it \textit{before} adding strain; in this case, the gap generated is a topologically nontrivial gap, accompanied by chiral edge states. The strain-induced field does not compete with this gap (because the lowest PLL is gapped out by the Haldane mass), thus this topologically trivial gap and the corresponding edge states must
continue to
exist even in strained graphene with a Haldane gap. On the other hand, the Haldane gap---when projected onto the zeroth PLL---can be rotated into the charge-density-wave gaps discussed in Ref.~\onlinecite{us:su2}; thus, its projection is topologically trivial and cannot generate edge states. Our numerical diagonalization studies, plotted in Fig.~\ref{zerothhaldane}, suggest that the resolution is as follows---the topologically protected Haldane edge state is connected to higher-order PLLs, and crosses the zeroth PLL without hybridizing with it. Thus, although these edge states give rise to a robust, quantized Hall effect, the mechanism is unrelated to the low-energy PLL structure. Indeed, in the limit where the Haldane mass term is much smaller than the other perturbations, we find that these edge states move far away from the Dirac points, demonstrating that they are unrelated to the PLLs.
In addition,
when the electron density is tuned to be at
the charge neutral point, and
the valley polarization (hence the Haldane mass)
is generated spontaneously by interactions, 
the interaction effects are
operative nominally only for the half-filled lowest PLL, but
not for the other PLLs that are totally empty or occupied. 
I.e., the Haldane mass is selectively
generated for the lowest PLL, but not for the other PLLs. 
We would then expect there is no edge state at all in this case.


\section{Conclusions and Outlook}\label{sec:conclusions}

In the present work we have addressed the properties of the edge states of strained graphene in the quantum Hall regime. We have argued that the edge physics is not universal---owing to the lack of topological protection---but is strongly dependent on the nature of the edge. Whereas armchair edges do not support edge states at all, zigzag edges are expected to support counter-propagating edge states. Furthermore, in the case of zigzag edges, the hybridization between the PLL states and the zigzag ``surface'' states gives rise to dispersion of the PLLs; for short-range interactions, these counter-propagating edges are expected to manifest themselves via nonmonotonic temperature-dependence of the conductance. Finally, we considered the case of valley-polarized edges (e.g., a quantum Hall ferromagnet in charge-neutral strained graphene) and argued that, in general, valley-polarization does not imply a finite Hall conductance, because (within the zeroth PLL) valley polarization can be continuously transformed into a charge-density wave, which is evidently non-topological.

Our results have several experimental implications. Most notably, the difference between the dispersion near zigzag and armchair edges can easily be detected using the spectroscopic methods of current experiments. In addition, transport experiments would---assuming the contribution of the PLLs could be isolated---provide several clear signatures. For example, for strained graphene ribbons having clean zigzag or armchair edges, the striking difference between the former case (with multiple counterpropagating edge modes) and the latter (no edge modes at all) should be easily detectable via two-terminal measurements. Similarly, a nonmonotonic temperature-dependence of the two-terminal conductance would provide a signature of interaction effects near the edges. Finally, either spectroscopy or transport can address our prediction that, for half-filled PLLs, the regions between magnetic domains should (a)~contain no propagating modes, and (b)~exhibit spontaneously broken translational symmetry. 


\section*{Acknowledgments}

The authors are grateful to Dmitry Abanin, Jeffrey Teo, Eduardo Fradkin, Taylor Hughes and Ashvin Vishwanath for helpful discussions. This work was supported by NSF
DMR-1064319 (P.G.) and DOE DE-FG02-07ER46453 (S.G.).

\bibliography{edgefinal}

\end{document}